\documentclass{article}
\usepackage{amssymb}

\usepackage{latexsym}
\usepackage{amsmath}
\usepackage{graphicx}
\usepackage{float}
\newlength{\dinwidth}
\newlength{\dinmargin}
\begin{document}

%--------------------------------------------------------------------------

%--------------------------------------------------------------------------
%--------------------------------------------------------------------------

\thispagestyle{empty} \vspace*{1cm}  %
 \vspace*{2cm}

\begin{center}
{\LARGE The m-reduction in Conformal Field Theory as the Morita equivalence
on two-tori }

{\LARGE \ }

{\large Vincenzo Marotta\footnote{{\large {\footnotesize Dipartimento di
Scienze Fisiche,}{\it \ {\footnotesize Universit\'{a} di Napoli ``Federico
II''\ \newline
and INFN, Sezione di Napoli}, }{\small Compl.\ universitario M. Sant'Angelo,
Via Cinthia, 80126 Napoli, Italy}}},} {\large Adele Naddeo\footnote{{\large
{\footnotesize CNISM, Unit\`{a} di Ricerca di Salerno and Dipartimento di
Fisica {\it ''}E. R. Caianiello'',}{\it \ {\footnotesize Universit\'{a}
degli Studi di Salerno, }}{\small Via Salvador Allende, 84081 Baronissi
(SA), Italy}}}}

{\small \ }

{\bf Abstract\\[0pt]
}
\end{center}

\begin{quotation}
We study the Morita equivalence for field theories on noncommutative
two-tori. For rational values of the noncommutativity parameter $\theta $
(in appropriate units) we show the equivalence between an abelian
noncommutative field theory and a nonabelian theory of twisted fields on
ordinary space. We concentrate on a particular conformal field theory (CFT),
the one obtained by means of the $m$-reduction procedure \cite{VM}, and show
that the Morita equivalence also holds at this level. An application to the
physics of a quantum Hall fluid at Jain fillings $\nu =\frac{m}{2pm+1}$ is
explicitly considered in order to further elucidate such a correspondence.

\vspace*{0.5cm}


{\footnotesize PACS: 11.25.Hf, 11.10.Nx, 03.65.Fd\newpage }\baselineskip%
=18pt \setcounter{page}{2}
\end{quotation}

\section{Introduction}

Noncommutative field theories (NCFT) have attracted much attention in the
last years because they provide a non trivial generalization of local
quantum field theories, allowing for some degree of non locality while
retaining an interesting mathematical structure \cite{ncft1}. Indeed they
naturally arise as some low energy limit of open string theory and as the
compactification of M-theory on the torus \cite{ncft2}. Space-time
noncommutativity also arises naturally when the dynamics of open strings
attached to a $D2$-brane in a $B$ field background is considered \cite{ncft3}%
: in such a case the open strings act as dipoles of $U\left( 1\right) $
gauge field of the brane and their scattering amplitudes in the low energy
limit are properly described by a Super Yang-Mills gauge theory defined on a
noncommutative two-torus with deformation parameter $\theta $ identified
with the\ $B$ field. More generally, gauge theories on tori with magnetic
flux and twisted models can be reformulated in terms of noncommutative gauge
theories. One motivation for the relevance of such theories is that the
notion of space-time presumably has to be modified at very short distances
so that the effect of the granularity of the space can be taken into account.

A significant feature of NCFT is the celebrated Morita duality between
noncommutative tori \cite{morita}. This duality is a powerful mathematical
result that establishes a relation, via an isomorphism, between two
noncommutative algebras. Of particular importance are the algebras defined
on the noncommutative torus, where it can be shown that Morita equivalence
holds if the corresponding sizes of the tori and the noncommutative
parameters are related in a specific way. Several results have been
established in the literature about the Morita equivalence of NCFT but
principally focused on noncommutative gauge theories and describing mostly
classical or semiclassical aspects of them \cite{morita1}. Indeed Morita
duality of gauge theories on noncommutative tori is a low energy analogue of
$T$-duality of the underlying string model \cite{string1}; when combined
with the hypothesis of analiticity as a function of the noncommutativity
parameter, it gives information about singular large-$N$ limits of ordinary $%
U(N)$ gauge theories \cite{alvarez1}. Nevertheless another point
of view can be usefully developed in order to establish a
correspondence between NCFT and well known standard field
theories. Indeed, for special values of the noncommutative
parameter, one of the isomorphic theories obtained by using the
Morita equivalence is a commutative field theory on an ordinary
space \cite{moreno1}.

Here we follow this line and concentrate on a particular conformal field
theory (CFT), the one obtained via $m$-reduction technique \cite{VM}, which
has been recently applied to the description of a quantum Hall fluid (QHF)
at Jain \cite{cgm1}\cite{cgm3} as well as paired states fillings \cite{cgm2}
\cite{cgm4} and in the presence of topological defects \cite{noi1}; by using
Morita duality we build up the essential ingredients of the corresponding
NCFT. In such a context the granularity of the space is due to the existence
of a minimum area which is the result of a fractionalized magnetic flux. The
$m$-reduction technique is based on the simple observation that, for any CFT
(mother), a class of sub-theories exists, which is parameterized by an
integer $m$ with the same symmetry but different representations. The
resulting theory (daughter), called Twisted Model (TM), has the same
algebraic structure but a different central charge $c_{m}=mc$. Its
application to the physics of the QHF arises by the incompressibility of the
Hall fluid droplet at the plateaux, which implies its invariance under the $%
W_{1+\infty }$ algebra at different fillings \cite{ctz5}, and by
the peculiarity of the $m$-reduction procedure to provide a
daughter CFT with the same $W_{1+\infty }$ invariance property of
the mother theory \cite{cgm1} \cite{cgm3}. Thus the $m$-reduction
furnishes automatically a mapping between different incompressible
plateaux of the QHF. The characteristics of the daughter theory is
the presence of twisted boundary conditions on the fundamental
fields, which coincide with the conditions required by the Morita
equivalence for a class of NCFT. Here the noncommutativity of the
spatial coordinates appears as a consequence of the twisting. As a
result, the $m$-reduction technique becomes the image in the
ordinary space of the Morita duality. Furthermore the Moyal
algebra, which characterizes the NCFT, has a natural realization
in terms of Generalized Magnetic Translations (GMT) within the
$m$-reduced theory when we refer to the description of a QHF at
Jain fillings $\nu =\frac{m}{2pm+1}$ \cite{cgm1}\cite{cgm3}. In
this paper we show how the $m$-reduction procedure induces a CFT
which is the Morita equivalent of a NCFT by making explicit
reference to the physics of a QHF at Jain fillings. We point out
that in the last years there have been many investigations on the
relationship between noncommutative spaces and
QHF: in all such studies noncommutativity is related to the finite number $%
N_{e}$ of electrons in a realistic sample via the rational parameter $\theta
\propto \frac{1}{N_{e}}$, which sets the elementary area of nonlocality \cite
{qhf1}\cite{qhf2}. In this context the $\theta $-dependence of physical
quantities is expected to be analytic near $\theta =0$ because of the very
smooth ultraviolet behaviour of noncommutative Chern-Simons theories \cite
{alvarez1}; as a consequence the effects of the electron's granularity
embodied in $\theta $ can be expanded in powers of $\theta $ via the
Seiberg-Witten map \cite{ncft2} and then appear as corrections to the large-$%
N$ results of field theory. Our approach is quite different: the
noncommutativity is present at field theory level and constrains the
structure of the theory already in the large-$N$ limit. It can be ascribed
to a residual noncommutativity for large numbers of electrons which group in
clusters of finite $m$.

The paper is organized as follows.

In Section 2, we review the main steps to be performed in order to get a $m$%
-reduced CFT on the plane \cite{VM} and then we briefly recall the
description of a QHF at Jain fillings $\nu =\frac{m}{2pm+1}$ as a result of
the $m$-reduction procedure \cite{cgm1}\cite{cgm3}.

In Section 3, we explicitly build up the Morita equivalence between CFTs in
correspondence of rational values of the noncommutativity parameter $\theta $
with an explicit reference to the $m$-reduced theory describing a QHF at
Jain fillings. Indeed we show that there is a well defined isomorphism
between the fields on a noncommutative torus and those of a non-abelian
field theory on an ordinary space.

In Section 4, we further clarify the deep relationship between $m$-reduction
procedure and Morita equivalence by showing how the NCFT properties can be
obtained from a generalization of the ordinary magnetic translations in a
QHF context \cite{cristofano1}.

In Section 5, some comments and outlooks are given.

\section{The $m$-reduction procedure}

In this Section we review the basics of the $m$-reduction
procedure on the plane (genus $g=0$) \cite{VM} and then we show
briefly how it works, referring to the description of a QHF at
Jain fillings $\nu =\frac{m}{2pm+1}$ \cite{cgm1}\cite{cgm3}.

In general, the $m$-reduction technique is based on the simple observation
that for any CFT (mother) exists a class of sub-theories parameterized by an
integer $m$ with the same symmetry but different representations. The
resulting theory (daughter) has the same algebraic structure but a different
central charge $c_{m}=mc$. In order to obtain the generators of the algebra
in the new theory we need to extract the modes which are divided by the
integer $m$. These can be used to reconstruct the primary fields of the
daughter CFT. This technique can be generalized and applied to any extended
chiral algebra which includes the Virasoro one. Following this line one can
generate a large class of CFTs with the same extended symmetry but with
different central extensions. It can be applied in particular to describe
the full class of Wess-Zumino-Witten (WZW) models with symmetry $\widehat{%
su(2)}_{m}$, obtaining the associated parafermions in a natural way or the
incompressible $W_{1+\infty }$ minimal models \cite{ctz5} with central
charge $c=m$. Indeed the $m$-reduction preserves the commutation relations
between the algebra generators but modifies the central extension (i.e. the
level for the WZW models). In particular this implies that the number of the
primary fields gets modified.

The general characteristics of the daughter theory is the presence of
twisted boundary conditions (TBC) which are induced on the component fields
and are the signature of an interaction with a localized topological defect.
It is illuminating to give a geometric interpretation of that in terms of
the covering on a $m$-sheeted surface or a complex curve with branch-cuts,
see for instance Fig. 1 for the particular case $m=2$.

\begin{figure}[h]
\centering\includegraphics*[width=0.3\linewidth]{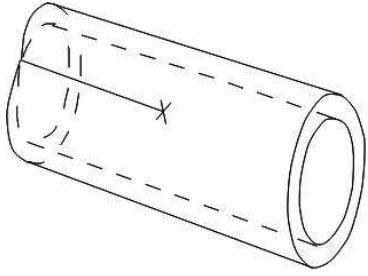}
\caption{The edge of the $2$-covered cylinder can be viewed as a separation
line of two different domains of the $2$-reduced CFT.}
\label{figura1}
\end{figure}

Indeed the fields which are defined on the left domain of the boundary have
TBC while the fields defined on the right one have periodic boundary
conditions (PBC). When we generalize the construction to a Riemann surface
of genus $g=1$, i. e. a torus, we find different sectors corresponding to
different boundary conditions on the cylinder, as shown in detail in Refs.
\cite{cgm3}\cite{cgm4}. Finally we recognize the daughter theory as an
orbifold of the usual CFT describing the QHF at a given plateau.

The physical interpretation of such a construction within the context of a
QHF description is the following. The two sheets simulate a two-layer
quantum Hall system and the branch cut represents TBC which emerge from the
interaction with a localized topological defect on the edge \cite{noi1}.

Let us now briefly summarize the $m$-reduction procedure on the plane \cite
{VM}, which has been recently applied to the description of a quantum Hall
fluid (QHF) at Jain \cite{cgm1} as well as non standard fillings \cite{cgm2}%
. Its generalization to the torus topology has been given in Refs. \cite
{cgm3}\cite{cgm4}. The starting point is described by a CFT with $c=1$, in
terms of a scalar chiral field compactified on a circle with general radius $%
R^{2}$ ($R^{2}=1$ for the Jain series \cite{cgm1} while $R^{2}=2$ for the
non standard one \cite{cgm2}). Then the $u(1)$ current is given by $%
J(z)=i\partial _{z}Q(z)$, where $Q(z)$ is the compactified Fubini field with
the standard mode expansion:
\begin{equation}
Q(z)=q-i\,p\,lnz+\sum_{n\neq 0}\frac{a_{n}}{n}z^{-n};
\label{modes}
\end{equation}
here $a_{n}$, $q$ and $p$ satisfy the commutation relations
$\left[
a_{n},a_{n^{\prime }}\right] =n\delta _{n,n^{\prime }}$ and $\left[ q,p%
\right] =i$. The primary fields are expressed in terms of the vertex
operators $U^{\alpha _{s}}(z)=:e^{i\alpha _{s}Q(z)}:$ with $\alpha _{s}=%
\frac{s}{R}$\ ($s=1,...,R^{2}$) and conformal dimension $h=\frac{s^{2}}{%
2R^{2}}$.

Starting with the set of fields in the above CFT and using the $m$-reduction
procedure we get the image of the twisted sector of a $c=m$ orbifold CFT (i.
e. the TM), which describes the Lowest Landau Level (LLL) dynamics of the
new filling in the QHF context. In this way the fundamental fields are
mapped into $m$ twisted fields which are related by a discrete abelian
group. Indeed the fields in the mother CFT can be factorized into
irreducible orbits of the discrete $Z_{m}$ group, which is a symmetry of the
TM, and can be organized into components, which have well defined
transformation properties under this group. To compare the orbifold so built
with the $c=m$ CFT, we use the mapping $z\rightarrow z^{1/m}$ and the
isomorphism, defined in Ref. \cite{VM}, between fields on the $z$ plane and
fields on the $z^{m}$ covering plane given by the following identifications:
$a_{nm+l}\longrightarrow \sqrt{m}a_{n+l/m}$, $q\longrightarrow \frac{1}{%
\sqrt{m}}q$.

We perform a \textquotedblleft double\textquotedblright\ $m$-reduction which
consists in applying this technique into two steps.

{\bf 1)} The $m$-reduction is applied to the Fubini field $Q(z).$ That
induces twisted boundary conditions on the currents. It is useful to define
the invariant scalar field:
\begin{equation}
X(z)=\frac{1}{m}\sum_{j=1}^{m}Q(\varepsilon ^{j}z),  \label{X}
\end{equation}
where $\varepsilon ^{j}=e^{i\frac{2\pi j}{m}}$, corresponding to a
compactified boson on a circle with radius now equal to $R_{X}^{2}=R^{2}/m$.
This field describes the $U(1)$ electrically charged component of the new
filling in a QHF description.

On the other hand the non-invariant fields defined by
\begin{equation}
\phi ^{j}(z)=Q(\varepsilon ^{j}z)-X(z),~~~~~~~~~~~~~~~~\sum_{j=1}^{m}\phi
^{j}(z)=0  \label{phi}
\end{equation}
naturally satisfy twisted boundary conditions, so that the $J(z)$ current of
the mother theory decomposes into a charged current given by $J(z)=i\partial
_{z}X(z)$ and $m-1$ neutral ones $\partial _{z}\phi ^{j}(z)$ \cite{cgm1}.

{\bf 2)} The $m$-reduction applied to the vertex operators $U^{\alpha
_{s}}(z)$ of the mother theory also induces twisted boundary conditions on
the vertex operators of the daughter CFT. The discrete group used in this
case is just the $m$-ality group which selects the neutral modes with a
complementary cut singularity, which is necessary to reinforce the locality
constraint.

The vertex operator in the mother theory can be factorized into a vertex
that depends only on the $X(z)$ field:
\begin{equation}
{\cal U}^{\alpha _{s}}(z)=z^{\frac{\alpha _{s}^{2}(m-1)}{m}}:e^{i\alpha _{s}{%
\ }X(z)}:
\end{equation}
and in vertex operators depending on the $\phi ^{j}(z)$ fields. It is useful
to introduce the neutral component:
\begin{equation}
\psi _{1}(z)=\frac{z^{\frac{1-m}{m}}}{m}\sum_{j=1}^{m}\varepsilon
^{j}:e^{i\phi ^{j}(z)}:
\end{equation}
which is invariant under the twist group given in {\bf 1)} and has $m$-ality
charge $l=1$. Then, the new primary fields are the composite vertex
operators $V^{\alpha _{s}}(z)={\cal U}^{\alpha _{s}}(z)\psi _{l}(z)$ where
{\bf $\psi _{l}$ }are the neutral operators with $m$-ality charge $l$.

\bigskip From these primary fields we can obtain the new Virasoro algebra
with central charge $c=m$ which is generated by the energy-momentum tensor $%
T(z)$. It is the sum of two independent operators, one depending on the
charged sector:
\begin{equation}
T_{X}(z)=-\frac{1}{2}:\left( \partial _{z}X\left( z\right) \right) ^{2}:
\label{VIR1}
\end{equation}
with $c=1$ and the other given in terms of the $Z_{m}$ twisted bosons $\phi
^{j}(z)$:
\begin{equation}
T_{\phi }(z)=-\frac{1}{2}\sum_{j,j^{\prime }=1}^{m}:\partial _{z}\phi
^{j}(z)\partial _{z}\phi ^{j^{\prime }}(z):+\ \frac{m^{2}-1}{24mz^{2}}
\label{VIR2}
\end{equation}
with $c=m-1$.

Let us notice here that, although the daughter CFT has the same central
charge value, it differs in the symmetry properties and in the spectrum,
depending on the mother theory we are considering (i.e. for Jain or non
standard series in the case of a QHF). Now, if we choose as starting point a
CFT with $c=1$, in terms of a scalar chiral field compactified on a circle
with radius $R^{2}=1$, see Eq. (\ref{modes}), we get the full spectrum of
excitations of a QHF at Jain fillings \cite{cgm1}. The dynamical symmetry is
given by the $W_{1+\infty }$ algebra \cite{BS} with $c=1$, whose generators
are simply given by a power of the current $J(z)$. Indeed, by using the $m$%
-reduction procedure, we get the image of the twisted sector of a $c=m$
orbifold CFT which has $\widehat{U}(1){\times }\widehat{SU}(m)_{1}$ as
extended symmetry and describes the QHF at the new general filling $\nu =%
\frac{m}{2pm+1}$. The above construction has been generalized to the torus
topology as well \cite{cgm3}, confirming the physical picture just outlined.

\section{Morita equivalence and the Twisted Model}

In this Section we explicitly build up the Morita equivalence between CFTs
in correspondence of some rational values of the noncommutativity parameter $%
\theta $ with an explicit reference to the $m$-reduced theory describing a
QHF at Jain fillings, recalled in Section 2.

The Morita equivalence \cite{morita}\cite{morita1} is an isomorphism between
noncommutative algebras that conserves all the modules and their associated
structures. Let us consider an $U(N)$ NCFT defined on the noncommutative
torus ${\rm T}_{\theta }^{2}$ and, for simplicity, of radii $R$. The
coordinates satisfy the commutation rule $[x_{1},x_{2}]=i\theta $ \cite
{ncft1}. In such a simple case the Morita duality is represented by the
following $SL(2,Z)$ action on the parameters:
\begin{equation}
\theta ^{^{\prime }}=\frac{a\theta +b}{c\theta +d};\\\\\\\
R^{^{\prime }}=\left| c\theta +d\right| R,  \label{morita1}
\end{equation}
where $a,b,c,d$ are integers and $ad-bc=1$.

For rational values of the non commutativity parameter, $\theta =-\frac{b}{a}
$, so that $c\theta +d=\frac{1}{a}$, the Morita transformation (\ref{morita1}%
) sends the NCFT to an ordinary one with $\theta ^{^{\prime }}=0$ and
different radius $R^{^{\prime }}=\frac{R}{a}$, involving in particular a
rescaling of the rank of the gauge group \cite{moreno1}\cite{alvarez1}.
Indeed the dual theory is a twisted $U(N^{^{\prime }})$ theory with $%
N^{^{\prime }}=aN$. The classes of $\theta ^{^{\prime }}=0$ theories are
parametrized by an integer $m$, so that for any $m$ there is a finite number
of abelian theories which are related by a subset of the transformations
given in Eq. (\ref{morita1}).

In this context the $m$-reduction technique applied to the QHF at Jain
fillings ($\nu =\frac{m}{2pm+1}$) can be viewed as the image of the Morita
map (characterized by $a=2p(m-1)+1$, $b=2p$, $c=m-1$, $d=1$) between the two
NCFTs with $\theta =1$ and $\theta =2p+\frac{1}{m}$ respectively and
corresponds to the Morita map in the ordinary space. Indeed the $\theta =1$
theory is an $U\left( 1\right) _{\theta =1}$ NCFT while the mother CFT is an
ordinary $U\left( 1\right) $ theory; furthermore, when we look at the $%
U\left( 1\right) _{\theta =2p+\frac{1}{m}}$ NCFT, its Morita dual CFT has $%
U\left( m\right) $ symmetry. The whole correspondence between the NCFTs and
the ordinary CFTs is summarized in the following table:
\begin{equation}
\begin{array}{ccc}
& \text{Morita} &  \\
U\left( 1\right) _{\theta =1} & \rightarrow & U\left( 1\right) _{\theta =0}
\\
& \left( a=1,b=-1,c=0,d=1\right) &  \\
\text{Morita}\downarrow \left( a,b,c,d\right) &  & m-\text{reduction}%
\downarrow \\
& \text{Morita} &  \\
U\left( 1\right) _{\theta =2p+\frac{1}{m}} & \rightarrow & U\left( m\right)
_{\theta =0} \\
& \left( a=m,b=-2pm-1,c=1-m,d=2p\left( m-1\right) +1\right) &
\end{array}
\label{morita2}
\end{equation}
It is the main result of this paper.

For more general commutativity parameters $\theta =\frac{q}{m}$ such a
correspondence can be easily extended. Indeed the action of the $m$%
-reduction procedure on the number $q$ doesn't change the central
charge of the CFT under study but modifies the compactification
radius of the charged sector \cite{cgm1}\cite{cgm3}. Nevertheless
in this paper we are interested to the action of the Morita map on
the denominator of the parameter $\theta $ which has interesting
consequences on noncommutativity, so in the following we will
concentrate on such an issue. The generalization of the Morita map
to different rational noncommutativity parameters, and at the same
time to the physics of QHF at different filling factors, will be
the subject of a future publication \cite{nos1}.

Let us now show in detail how the twisted boundary conditions on the neutral
fields of the $m$-reduced theory (see Section 2) arise as a consequence of
the noncommutative nature of the $U\left( 1\right) _{\theta =2p+\frac{1}{m}}$
NCFT. In order to carry out this program let us recall that an associative
algebra of smooth functions over the noncommutative two-torus ${\rm T}%
_{\theta }^{2}$ can be realized through the Moyal product ($%
[x_{1},x_{2}]=i\theta $):
\begin{equation}
f\left( x\right) \ast g\left( x\right) =\left. \exp \left( \frac{i\theta }{2}%
\left( \partial _{x_{1}}\partial _{y_{2}}-\partial _{x_{2}}\partial
_{y_{1}}\right) \right) f\left( x\right) \ast g\left( y\right) \right\vert
_{y=x}.  \label{moyal1}
\end{equation}
It is convenient to decompose the elements of the algebra, i. e. the fields,
in their Fourier components. However a general field operator $\Phi $
defined on a torus can have different boundary conditions associated to any
of the compact directions. For the torus we have four different
possibilities:
\begin{equation}
\begin{array}{cc}
\Phi \left( x_{1}+R,x_{2}\right) =e^{2\pi i\alpha _{1}}\Phi \left(
x_{1},x_{2}\right) , & \Phi \left( x_{1},x_{2}+R\right) =e^{2\pi i\alpha
_{2}}\Phi \left( x_{1},x_{2}\right) ,
\end{array}
\label{bc1}
\end{equation}
where $\alpha _{1}$ and $\alpha _{2}$ are the boundary parameters. The
Fourier expansion of the general field operator $\Phi _{\overrightarrow{%
\alpha }}$ with boundary conditions $\overrightarrow{\alpha }=\left( \alpha
_{1},\alpha _{2}\right) $ takes the form:
\begin{equation}
\Phi _{\overrightarrow{\alpha }}=\sum_{\overrightarrow{n}}\Phi ^{%
\overrightarrow{n}}U_{\overrightarrow{n}+\overrightarrow{\alpha }}
\label{fexp1}
\end{equation}
where we define the generators as
\begin{equation}
U_{\overrightarrow{n}}\equiv \exp \left( 2\pi i\frac{\overrightarrow{n}\cdot
\overrightarrow{x}}{R}\right) .  \label{fexp2}
\end{equation}
They give rise to the following Moyal commutator:
\begin{equation}
\left[ U_{\overrightarrow{n}+\overrightarrow{\alpha }},U_{\overrightarrow{%
n^{\prime }}+\overrightarrow{\alpha ^{\prime }}}\right] =-2i\sin \left(
\frac{2\pi ^{2}\theta }{R^{2}}\left( \overrightarrow{n}+\overrightarrow{%
\alpha }\right) \wedge \left( \overrightarrow{n^{\prime }}+\overrightarrow{%
\alpha ^{\prime }}\right) \right) U_{\overrightarrow{n}+\overrightarrow{%
n^{\prime }}+\overrightarrow{\alpha }+\overrightarrow{\alpha ^{\prime }}},
\label{fexp3}
\end{equation}
where $\overrightarrow{p}\wedge \overrightarrow{q}=\varepsilon
_{ij}p_{i}q_{j}$.

When the noncommutativity parameter $\theta $ takes the rational value $%
\theta =\frac{2q}{m}\frac{R^{2}}{2\pi }$, being $q$ and $m$ relatively prime
integers, the infinite-dimensional algebra generated by the $U_{%
\overrightarrow{n}+\overrightarrow{\alpha }}$ breaks up into
equivalence classes of finite dimensional subspaces. Indeed the
elements $U_{m\overrightarrow{n}}$ generate the center of the
algebra and that makes possible for the momenta the following
decomposition:
\begin{equation}
\overrightarrow{n^{\prime }}+\overrightarrow{\alpha }=m\overrightarrow{n}+%
\overrightarrow{n},\\\\\\\ 0\leq n_{1},n_{2}\leq m-1.
\label{fexp4}
\end{equation}
The whole algebra splits into equivalence classes classified by all the
possible values of $m\overrightarrow{n}$, each class being a subalgebra
generated by the $m^{2}$ functions $U_{\overrightarrow{n}+\overrightarrow{%
\alpha }}$ which satisfy the relations
\begin{equation}
\left[ U_{\overrightarrow{n}+\overrightarrow{\alpha }},U_{\overrightarrow{%
n^{\prime }}+\overrightarrow{\alpha ^{\prime }}}\right] =-2i\sin \left(
\frac{2\pi q}{m}\left( \overrightarrow{n}+\overrightarrow{\alpha }\right)
\wedge \left( \overrightarrow{n^{\prime }}+\overrightarrow{\alpha ^{\prime }}%
\right) \right) U_{\overrightarrow{n}+\overrightarrow{n^{\prime }}+%
\overrightarrow{\alpha }+\overrightarrow{\alpha ^{\prime }}}.  \label{fexp5}
\end{equation}
The algebra (\ref{fexp5}) is isomorphic to the (complexification of the) $%
U\left( m\right) $ algebra, whose general $m$-dimensional representation can
be constructed by means of the following ''shift''\ and ''clock''\ matrices
\cite{matrix2}:
\begin{equation}
Q=\left(
\begin{array}{cccc}
1 &  &  &  \\
& \varepsilon &  &  \\
&  & \ddots &  \\
&  &  & \varepsilon ^{m-1}
\end{array}
\right) ,\\\\\\\ P=\left(
\begin{array}{cccc}
0 & 1 &  & 0 \\
& \cdots &  &  \\
&  & \vdots & 1 \\
1 &  &  & 0
\end{array}
\right) ,  \label{fexp6}
\end{equation}
being $\varepsilon =\exp (\frac{2\pi iq}{m})$. So the matrices $J_{%
\overrightarrow{n}}=\varepsilon ^{n_{1}n_{2}}Q^{n_{1}}P^{n_{2}}$, $%
n_{1},n_{2}=0,...,m-1$, generate an algebra isomorphic to (\ref{fexp5}):
\begin{equation}
\left[ J_{\overrightarrow{n}},J_{\overrightarrow{n^{\prime }}}\right]
=-2i\sin \left( 2\pi \frac{q}{m}\overrightarrow{n}\wedge \overrightarrow{%
n^{\prime }}\right) J_{\overrightarrow{n}+\overrightarrow{n^{\prime }}}.
\label{fexp7}
\end{equation}
Thus the following Morita mapping has been realized between the Fourier
modes defined on a noncommutative torus and functions taking values on $%
U\left( m\right) $ but defined on a commutative space:
\begin{equation}
\exp \left( 2\pi i\frac{\left( \overrightarrow{n}+\overrightarrow{\alpha }%
\right) \cdot \widehat{\overrightarrow{x}}}{R}\right) \longleftrightarrow
\exp \left( 2\pi i\frac{\left( \overrightarrow{n}+\overrightarrow{\alpha }%
\right) \cdot \overrightarrow{x}}{R}\right) J_{\overrightarrow{n}+%
\overrightarrow{\alpha }}.  \label{fexp8}
\end{equation}
As a consequence a mapping between the fields $\Phi _{\overrightarrow{\alpha
}}$ is generated as follows. Let us focus, for simplicity, on the case $q=1$
which leads for the momenta to the decomposition $\overrightarrow{n}=m%
\overrightarrow{n}+\overrightarrow{j}$, with \ $0\leq j_{1},j_{2}\leq m$.
The general field operator $\Phi _{\overrightarrow{\alpha }}$ on the
noncommutative torus ${\rm T}_{\theta }^{2}$ with boundary conditions $%
\overrightarrow{\alpha }$ can be written in the form:
\begin{equation}
\Phi _{\overrightarrow{\alpha }}=\sum_{\overrightarrow{q}}\exp \left( 2\pi im%
\frac{\overrightarrow{n}\cdot \overrightarrow{x}}{R}\right) \sum_{%
\overrightarrow{j}=0}^{m-1}\Phi ^{\overrightarrow{n},\overrightarrow{j}}U_{%
\overrightarrow{j}+\overrightarrow{\alpha }}.  \label{fexp9}
\end{equation}
By using Eq. (\ref{fexp8}) we obtain the Morita correspondence between
fields as:
\begin{equation}
\Phi _{\overrightarrow{\alpha }}\longleftrightarrow \Phi =\sum_{%
\overrightarrow{j}=0}^{m-1}\chi ^{\left( \overrightarrow{j}\right) }J_{%
\overrightarrow{j}+\overrightarrow{\alpha }},  \label{fexp10}
\end{equation}
where we have defined:
\begin{equation}
\chi ^{\left( \overrightarrow{j}\right) }=\exp \left( 2\pi i\frac{\left(
\overrightarrow{j}+\overrightarrow{\alpha }\right) \cdot \overrightarrow{x}}{%
R}\right) \sum_{\overrightarrow{q}}\Phi ^{\overrightarrow{n},\overrightarrow{%
j}}\exp \left( 2\pi im\frac{\overrightarrow{n}\cdot \overrightarrow{x}}{R}%
\right) .  \label{fexp11}
\end{equation}
The field $\Phi $ is defined on the dual torus with radius $R^{\prime }=%
\frac{R}{m}$ and satisfies the boundary conditions:
\begin{equation}
\begin{array}{cc}
\Phi \left( \theta +R^{\prime },x_{2}\right) =\Omega _{1}^{+}\cdot \Phi
\left( \theta ,x_{2}\right) \cdot \Omega _{1}, & \Phi \left( \theta
,x_{2}+R^{\prime }\right) =\Omega _{2}^{+}\cdot \Phi \left( \theta
,x_{2}\right) \cdot \Omega _{2},
\end{array}
\label{fexp12}
\end{equation}
with
\begin{equation}
\Omega _{1}=P^{b},\\\\\\\ \Omega _{2}=Q^{1/q},  \label{fexp13}
\end{equation}
where $b$ is an integer satisfying $am-bq=1$. While the field components $%
\chi ^{\left( \overrightarrow{j}\right) }$ satisfy the following twisted
boundary conditions:
\begin{equation}
\begin{array}{c}
\chi ^{\left( \overrightarrow{j}\right) }\left( \theta +R^{\prime
},x_{2}\right) =e^{2\pi i\left( j_{1}+\alpha _{1}\right) /m}\chi ^{\left(
\overrightarrow{j}\right) }\left( \theta ,x_{2}\right) \\
\chi ^{\left( \overrightarrow{j}\right) }\left( \theta ,x_{2}+R^{\prime
}\right) =e^{2\pi i\left( j_{2}+\alpha _{2}\right) /m}\chi ^{\left(
\overrightarrow{j}\right) }\left( \theta ,x_{2}\right)
\end{array}
,  \label{fexp14}
\end{equation}
that is
\begin{equation}
\left( \frac{j_{1}+\alpha _{1}}{m},\frac{j_{2}+\alpha
_{2}}{m}\right) ;\\\\\\\ j_{1}=0,...,m-1;\\\\\\\ j_{2}=0,...,m-1.
\label{fexp15}
\end{equation}
Let us observe that $\overrightarrow{j}=\left( 0,0\right) $ is the trace
degree of freedom which can be identified with the $U(1)$ component of the
matrix valued field or the charged component within the $m$-reduced theory
of the QHF at Jain fillings introduced in Section 2. We conclude that only
the integer part of $\frac{n_{i}}{m}$ should really be thought of as the
momentum. The commutative torus is smaller by a factor $m\times m$ than the
noncommutative one; in fact upon this rescaling also the ''density of
degrees of freedom''\ is kept constant as now we are dealing with $m\times m$
matrices instead of scalars.

Summarizing, when the parameter $\theta $ is rational we recover the whole
structure of the noncommutative torus and recognize the twisted boundary
conditions which characterize the neutral fields (\ref{phi}) of the $m$%
-reduced theory as the consequence of the Morita mapping of the
starting NCFT ($U\left( 1\right) _{\theta =2p+\frac{1}{m}}$ in our
case) on the ordinary commutative space. Indeed $\chi ^{\left(
0,0\right) }$\ corresponds to the charged $X$\ field while the
twisted fields $\chi ^{\left(\overrightarrow{j}\right) }$ with $\overrightarrow{j}\neq \left( 0,0\right) $%
\ should be identified with the neutral ones (\ref{phi}). Therefore the $m$%
-reduction technique can be viewed as a realization of the Morita mapping
between NCFTs and CFTs on the ordinary space, as sketched in the table (\ref
{morita2}). In the next Section we further clarify such a correspondence by
making an explicit reference to the QHF physics at Jain fillings. In
particular we will recognize the GMT as a realization of the Moyal algebra
defined in Eq. (\ref{fexp5}).

\section{Moyal algebra and generalized magnetic translations}

In this Section we will make explicit reference to the $m$-reduced theory
for a QHF at Jain fillings and to the issue of GMT on a torus in order to
identify in such a context a realization of the Moyal algebra.

Let us recall that all the relevant topological effects in the QHF
physics, such as the degeneracy of the ground state wave function
on a manifold with non trivial topology, the derivation of the
Hall conductance $\sigma _{H}$ as a topological invariant and the
relation between fractional charge and statistics of anyon
excitations \cite{top1}\cite{cristofano1}, can be made very
transparent by using the invariance properties of the wave
functions under a finite subgroup of the magnetic translation
group for a $N_{e}$ electrons system. Indeed their explicit
expressions as the Verlinde operators \cite{cristofano1} which
generate the modular transformations in the $c=1$ CFT are taken as
a realization of topological order of the system under study \cite
{wen}. In particular the magnetic translations built so far \cite
{cristofano1} act on the characters associated to the highest
weight states which represent the charged statistical particles,
the anyons or the electrons.

In our CFT representation of the QHF at Jain fillings \cite{cgm1}\cite{cgm3}
we shall see that the Moyal algebra defined in Eq. (\ref{fexp5}) has a
natural and beautiful realization in terms of GMT. We refer to them as
generalized ones because the usual magnetic translations act on the charged
content of the one point functions \cite{cristofano1}. Instead, in our TM
model for the QHF (see Section 2 and Refs. \cite{cgm3}, \cite{cgm4}) the
primary fields (and then the corresponding characters within the torus
topology) appear as composite field operators which factorize in a charged
as well as a neutral part. Further they are also coupled by the discrete
symmetry group $Z_{m}$. Then, in order to show that the characters of the
theory are closed under magnetic translations, we need to generalize them in
such a way that they will appear as operators with two factors, acting on
the charged and on the neutral sector respectively.

Let us also recall that the incompressibility of the quantum Hall
fluid naturally leads to a $W_{1+\infty }$ dynamical symmetry
\cite{BS, ctz}. Indeed, if one considers a droplet of a quantum
Hall fluid, it is evident that the only possible area preserving
deformations of this droplet are the waves at the boundary of the
droplet, which describe the deformations of its shape, the so
called edge excitations. These can be well described by the
infinite generators $W_{m}^{n+1}$ of $W_{1+\infty }$ of conformal spin ($n+1$%
), which are characterized by a mode index $m\in Z$ and satisfy the algebra:
\begin{equation}
\left[ W_{m}^{n+1},W_{m^{\prime }}^{n^{\prime }+1}\right] =(n^{\prime
}m-nm^{\prime })W_{m+m^{\prime }}^{n+n^{\prime }}+q(n,n^{\prime
},m,m^{\prime })W_{m+m^{\prime }}^{n+n^{\prime }-2}+...+d(n,m)c\,\delta
^{n,n^{\prime }}\delta _{m+m^{\prime }=0}  \label{walgebras}
\end{equation}
where the structure constants $q$ and $d$ are polynomials of their
arguments, $c$ is the central charge, and dots denote a finite number of
similar terms involving the operators $W_{m+m^{\prime }}^{n+n^{\prime }-2l}$
\cite{BS, ctz}. Such an algebra contains an Abelian $\widehat{U}(1)$ current
for $n=0$ and a Virasoro algebra for $n=1$ with central charge $c$. It
encodes the local properties which are imposed by the incompressibility
constraint and realizes the allowed edge excitations \cite{ctz}.
Nevertheless algebraic properties do not include topological properties
which are also a consequence of incompressibility. In order to take into
account the topological properties we have to resort to finite magnetic
translations which encode the large scale behaviour of the QHF.

Let us consider a magnetic translation of step $(n=n_{1}+in_{2},%
\overline{n}=n_{1}-in_{2})$ on a sample with coordinates $%
x_{1},x_{2}$ and define the corresponding generators $T^{n,\overline{n}}$
as:
\begin{equation}
T^{n,\overline{n}}=e^{-\frac{B}{4}n\overline{n}}e^{\frac{1}{2}nb^{+}}e^{%
\frac{1}{2}\overline{n}b},  \label{tr1}
\end{equation}
where $b^{+}=i\partial _{\overline{\omega }}-i\frac{B}{2}\omega $, $%
b=i\partial _{\omega }+i\frac{B}{2}\overline{\omega }$, $\omega $ is a
complex coordinate ($\overline{\omega }$ being its conjugate) and $B$ is the
transverse magnetic field. They satisfy the relevant property:
\begin{equation}
T^{n,\overline{n}}T^{m,\overline{m}}=q^{-\frac{n\times m}{4}}T^{n+m,}{}^{%
\overline{n}+\overline{m}},  \label{tr2}
\end{equation}
where $q$ is a root of unity.

Furthermore, it can be easily shown that they admit the following expansion
in terms of the generators $W_{k-1}^{l-1}$ of the $W_{1+\infty }$ algebra:
\begin{equation}
T^{n,\overline{n}}=e^{-\frac{B}{4}n\overline{n}}\sum_{k,l=0}^{\infty }\left(
-\right) ^{l}\frac{n^{k}}{2^{k}n}\frac{\overline{n}^{l}}{2^{l}\overline{n}}%
W_{k-1}^{l-1},  \label{tr3}
\end{equation}
where now the local $W_{1+\infty }$ symmetry and the global
topological properties are much more evident because the
coefficients in the above series depend on the topology of the
sample.

Within our $m$-reduced theory for a QHF at Jain fillings
\cite{cgm1}\cite {cgm3} it can be shown that also magnetic
translations of step ($n,\overline{n}$) decompose in equivalence
classes and can be factorized into a group, with generators
$T_{C}^{n,\overline{n}}$, which acts only on the charged sector as
well as a group, with generators $T_{S}^{j_{i},\overline{j}}$,
acting only on the neutral sector \cite{nos2}. The presence of the
transverse magnetic field $B$ reduces the torus to a
noncommutative one and the flux
quantization induces rational values of the noncommutativity parameter $%
\theta $. As a consequence the neutral magnetic translations realize a
projective representation of the $su\left( m\right) $ algebra generated by
the elementary translations:
\begin{equation}
J_{a,b}=e^{-2\pi i\frac{ab}{m}}T_{S}^{a,0}T_{S}^{0,b};\\\\\\\
a,b=1,...,m,  \label{tr4}
\end{equation}
which satisfy the commutation relations:
\begin{equation}
\left[ J_{a,b},J_{\alpha ,\beta }\right] =-2i\sin \left( \frac{2\pi }{m}%
\left( a\beta -b\alpha \right) \right) J_{a+\alpha ,b+\beta }.  \label{tr5}
\end{equation}
The GMT operators above defined (see Eqs. (\ref{tr1}) and
(\ref{tr4})) are a realization of the Moyal operators introduced
in Eq. (\ref{fexp2}) and the algebra defined by Eq. (\ref{tr5}) is
isomorphic to the Moyal algebra given in Eqs. (\ref{fexp5}) and
(\ref{fexp7}). Such operators generate the residual symmetry of
the $m$-reduced CFT which is Morita equivalent to the NCFT with
rational non commutativity parameter $\theta =2p+\frac{1}{m}$.

\section{Conclusions and outlooks}

In conclusion, the Morita equivalence gives rise to an isomorphism
between noncommutative algebras but, in the special case of a
rational noncommutativity parameter $\theta $, one of the
isomorphic theories is a commutative one \cite{moreno1}. In this
paper we have shown by means of the Morita equivalence that a NCFT
with $\theta =2p+\frac{1}{m}$ is mapped to a CFT on an ordinary
space. We identified such a CFT with the $m$-reduced CFT developed
in \cite{cgm1}\cite{cgm3} for a QHF at Jain fillings, whose
neutral fields satisfy twisted boundary conditions. In this way we
gave a meaning to the concept of ''noncommutative conformal field
theory'', as the Morita equivalent version of a CFT defined on an
ordinary space. The image of Morita duality in the ordinary space
is given by the $m$-reduction technique and the Moyal algebra
which reflects the noncommutative nature is realized by means of
GMT. Furthermore a new relation between noncommutative spaces and
QHF physics has been established within a perspective which is
very different from the ones developed in the literature in the
last years \cite{qhf1}\cite{qhf2}. The generalization of the
Morita map to different rational noncommutativity parameters
$\theta $, and at the same time to the physics of QHF at paired
state filling factors, will be the subject of a future publication
\cite{nos1}. That will help us to shed new light also on the
connections between a system of interacting $D$-branes and the
physics of quantum Hall fluids \cite{branehall1}. Recently the
twisted CFT approach provided by $m$-reduction has been
successfully applied to Josephson junction ladders and arrays of
non trivial geometry in order to investigate the existence of
topological order and magnetic flux fractionalization in view of
the implementation of a possible solid state qubit protected from
decoherence \cite{noi3} as well as to the study of the phase
diagram of the fully frustrated $XY$ model on a square lattice
\cite{noi}. So, it could be interesting to further elucidate the
topological properties of Josephson systems with non trivial
geometry and the consequences imposed by the request of space-time
noncommutativity by means of Morita mapping. Finally, the
generality of the $m$-reduction technique allows us to extend many
of the present results to $D>2$ theories as string and $M$-theory.

\end{document}